\documentclass[12pt]{article}
\usepackage{amssymb,amsmath,amsbsy}
\usepackage{dsfont}
\usepackage{cite}
\usepackage{mathrsfs}
\usepackage[title]{appendix}
\usepackage[hypertexnames=false,hyperfootnotes=false,pdfencoding =auto]{hyperref}
\makeatletter
\g@addto@macro\bfseries{\boldmath}
\makeatother

\DeclareMathOperator{\Tr}{Tr}

\def\ifmath#1{\relax\ifmmode #1\else $#1$\fi}
\def\half{\tfrac{1}{2}}
\def\ls#1{\ifmath{_{\lower1.5pt\hbox{$\scriptstyle #1$}}}}
\def\lsup#1{\ifmath{^{\lower3pt\hbox{$\scriptstyle #1$}}}}

\newenvironment{Eqnarray}%
     {\arraycolsep 0.14em\begin{eqnarray}}{\end{eqnarray}}
\newcommand{\beqa}{\begin{Eqnarray}}
\newcommand{\eeqa}{\end{Eqnarray}}

\def\beq{\begin{equation}}
\def\eeq{\end{equation}}
\def\eq#1{eq.~(\ref{#1})}
\def\Eq#1{Eq.~(\ref{#1})}

\def\eqs#1#2{eqs.~(\ref{#1}) and (\ref{#2})}
\def\eqss#1#2#3{eqs.~(\ref{#1}), (\ref{#2}) and (\ref{#3})}

\def\eqst#1#2{eqs.~(\ref{#1})--(\ref{#2})}

\def\id{{\bf I}}

\def\newcdot{\kern.06em{\cdot}\kern.06em}
\def\half{\tfrac{1}{2}}
\def\lsub#1{_{\lower 1.5pt\hbox{$\scriptstyle#1$}}}

\begin{document}
\begin{center}
{\Large \bf  Useful relations among the generators in the defining and adjoint representations of SU($N$)}\\[1cm]
{\Large Howard E. Haber}\\[5pt]
{\large Santa Cruz Institute for Particle Physics  \\[4pt]
   University of California, Santa Cruz, CA 95064, USA} \\[5pt]
\end{center}

\begin{abstract}

There are numerous relations among the generators in the defining and adjoint representations of  SU($N$).  
These include Casimir operators, formulae for traces of products of generators, etc.  Due to the existence
of the completely symmetric tensor $d_{abc}$ that arises in the study of the SU($N$) Lie algebra, one can also consider relations
that involve the adjoint representation matrix, $(D^a)_{bc}=d_{abc}$.  In this review, we summarize many useful relations satisfied
by the defining and adjoint representation matrices of SU($N$).  A few relations special to the case of $N=3$ are highlighted.
\end{abstract}

\section{Introduction}
The SU($N$) Lie group and its Lie algebra are ubiquitous in theoretical physics.  
Numerous relations among the generators in the defining and adjoint representations of  SU($N$) are often useful in a variety of physics applications.  
Many of these relations are well known and others are more obscure.  There are multiple sources for the various identities
that will be reviewed in these notes, but there is no single reference that I am aware of that contains all of them.
For my own benefit, as well as for the benefit of others, I have collected many of the relevant identities and assembled them in this 
short review.  In a few instances, I found typographical errors in some of the original sources that I was able to correct.
 
Recently, two authors contacted me concerning a first draft of these notes, which they apparently found to be quite useful~\cite{Arnold:2019qqc,private}.  They urged me to 
make these notes more widely available, and I am pleased to accommodate their request.   Although I have not taken the opportunity for providing
a more comprehensive list of references,  I have included references to the primary sources that were used in obtaining all the formulae of this review.

\section{The defining representation of the SU($N$) Lie algebra}

In these notes, we provide some useful relations involving the generators of the SU($N$) Lie algebra, henceforth denoted by
$\mathfrak{su}(N)$.   We employ the physicist's convention, where the  $N^2-1$ generators in the defining representation of $\mathfrak{su}(N)$, denoted by~$T^a$, serve as a basis for
the set of traceless hermitian $N\times N$ matrices.   The generators satisfy the commutation relations,
\beq \label{commutator}
[T^a\,,\,T^b]=if_{abc} T^c\,,\qquad\quad \text{where $a,b,c=1,2,\ldots,N^2-1$.}
\eeq
In particular
\beq \label{traceless}
\Tr T^a=0\,.
\eeq
We employ the following normalization convention for the generators in the defining representation of $\mathfrak{su}(N)$,
\beq \label{trtt}
\Tr (T^a T^b)=\half\delta_{ab}\,.
\eeq
In this convention, the $f^{abc}$ are totally antisymmetric with respect to the interchange of any pair of its indices.

Consider a $d$-dimensional irreducible representation, $R^a$ of the generators of $\mathfrak{su}(N)$.
The quadratic Casimir operator, $C_2\equiv R^a R^a$, commutes with all the $\mathfrak{su}(N)$ generators.\footnote{It is straightforward to show that $C_2$ commutes with all the generators of $\mathfrak{su}(N)$.  In particular, using the commutation relations, $[R^a\,,\,R^b]=if_{abc} R^c$,
$$
\bigl[R^a R^a\,,\,R^b\bigr]=R^a\bigl[R^a\,,\,R^b\bigr]+\bigl[R^a\,,\,R^b\bigr]R^a=if^{abc}(R^a R^c+R^c R^a)=0\,,
$$
due to the antisymmetry of $f^{abc}$ under the interchange of any pair of indices.}  Hence in light of Schur's lemma, $C_2$ is proportional to the $d\times d$ identity matrix.  In particular, the quadratic Casimir operator in the defining representation of $\mathfrak{su}(N)$ is given by
\beq \label{casimir}
T^a T^a=C_F\mathds{1}\,,
\eeq
where $\mathds{1}$ is the $N\times N$ identity matrix.
To evaluate $C_F$, we take the trace of \eq{casimir} and make use of $\Tr\mathds{1}=N$.  Summing over $a$, we note that $\delta_{aa}=N^2-1$.  Using the normalization of the generators specified in \eq{trtt}, it follows that $\half(N^2-1)=NC_F$.  Hence,\footnote{In the older literature, the defining representation is (inaccurately) called the fundamental representation.  It is for this reason that the Casimir operator in the defining representation is often denoted by $C_F$.}
\beq \label{cf}
C_F=\frac{N^2-1}{2N}\,.
\eeq

Next we quote an important identity involving the $\mathfrak{su}(N)$ generators in the defining representation,
\beq \label{tata}
T^a_{ij}T^a_{k\ell}=\frac12\left(\delta_{i\ell}\delta_{jk}-\frac{1}{N}\delta_{ij}\delta_{k\ell}\right)\,,
\eeq
where the indices $i$, $j$, $k$ and $\ell$ take on values from $1,2,\ldots,N$.
To derive \eq{tata}, we first note that any $N\times N$ complex matrix $M$ can be written as a complex linear combination of the $N\times N$ identity matrix and the $T^a$,
\beq \label{M}
M=M_0\mathds{1}+M_a T^a\,.
\eeq
This can be regarded as a completeness relation on the vector space of complex $N\times N$ matrices.
One can project out the coefficient $M_0$ by taking the trace of \eq{M}.  Likewise, one can project out the coefficients $M_a$ by multiplying \eq{M} by $T^b$ and then taking the trace of the resulting equation.
Using \eqs{traceless}{trtt}, it follows that
\beq
M_0=\frac{1}{N}\Tr M\,,\qquad\quad M_a=2\Tr(MT^a)\,.
\eeq
Inserting these results back into \eq{M} yields
\beq \label{M2}
M=\frac{1}{N}(\Tr M)\mathds{1}+2\Tr(MT^a)T^a\,.
\eeq 
The matrix elements of \eq{M2} are therefore
\beq \label{M3}
M_{ij}=\frac{1}{N} M_{kk}\delta_{ij}+2M_{\ell k} T^a_{k\ell} T^a_{ij}\,,
\eeq
where the sum over repeated indices is implicit.  We can rewrite \eq{M3} in a more useful form,
\beq \label{M4}
\delta_{i\ell}\delta_{jk}M_{\ell k}=\left(\frac{1}{N}\delta_{ij}\delta_{k\ell}+2 T^a_{ij}T^a_{k\ell} \right)M_{\ell k}\,.
\eeq
It follows that
\beq \label{M5}
\left[T^a_{ij}T^a_{k\ell} -\frac12\left(\delta_{i\ell}\delta_{jk}-\frac{1}{N}\delta_{ij}\delta_{k\ell}\right)\right]M_{\ell k}=0\,.
\eeq
This equation must be true for any arbitrary $N\times N$ complex matrix $M$.   It follows that the coefficient of $M_{\ell k}$ in \eq{M5} must vanish.  This yields the identity states in \eq{tata}.  The proof is complete.

Many important identities can be obtained from \eq{tata}.  For example, multiplying \eq{tata} by $T^b_{jk}$ and summing over $j$ and $k$ yields
\beq \label{tatbta}
T^a T^b T^a=-\frac{1}{2N}T^a\,,
\eeq
after employing \eq{traceless}.   If we now multiply \eq{tatbta} by $T^c$ and take the trace of both sides of the resulting equation, then the end result is
\beq \label{tatbtatc}
\Tr(T^a T^b T^a T^c)=-\frac{1}{4N}\delta_{bc}\,.
\eeq
after using \eq{trtt}.   A more general expression for the trace of four generators (of which \eq{tatbtatc} is a special case) is given in Appendix A.

\section{Introducing the symmetric third rank tensor $d_{abc}$}

In $\mathfrak{su}(N)$, one can also define a totally symmetric third rank tensor called $d^{abc}$ via the relation,
\beq \label{tatb}
T^a T^b=\frac12\left[\frac{1}{N}\delta_{ab}\mathds{1}+(d_{abc}+if_{abc})T^c\right]\,,
\eeq
where $\mathds{1}$ is the $N\times N$ identity matrix.  Combining \eqs{commutator}{tatb} yields the following anticommutation relation,
\beq \label{anticomm}
\bigl\{T^a\,,\,T^b\bigr\}\equiv T^a T^b+T^b T^a=\frac{1}{N}\delta_{ab}\mathds{1}+d_{abc} T^c\,,
\eeq
Using \eqs{trtt}{anticomm}, one obtains an explicit expression,
\beq
\label{dabc}
d_{abc}=2\Tr\bigl[\bigl\{T^a\,,\,T^b\bigr\}T^c\bigr]\,,
\eeq
which can be taken as the definition of the $d_{abc}$.  It then follows that $d_{aac}=0$ (where a sum over the repeated index $a$ is implicit).
Indeed, since $d_{abc}$ is a totally symmetric tensor, it follows that $d_{aca}=d_{caa}=0$.

The case of $\mathfrak{su}(2)$ provides the simplest example.  In this case, we identify $T^a=\half\sigma^a$, where the $\sigma^a$ (for $a=1,2,3$) are the well-known Pauli matrices, and $f_{abc}=\epsilon_{abc}$ are the components of the Levi-Civita tensor.  It is a simple matter to check that in the case of $\mathfrak{su}(2)$, we have $d_{abc}=0$.   In contrast, the $d_{abc}$ are generally non-zero for $N\geq 3$.

Consider the trace identity obtained by multiplying \eq{tatb} by $T^c$ and taking the trace.   In light of  \eqs{traceless}{trtt},
\beq \label{ttt}
\Tr(T^a T^b T^c)=\tfrac{1}{4}\left(d_{abc}+if_{abc}\right)\,.
\eeq
It then follows that
\beqa
f_{abd}\Tr(T^a T^b T^c)&=&\tfrac{1}{4}if_{abc}f_{abd}\,,\label{fttt}\\[6pt]
d_{abd}\Tr(T^a T^b T^c)&=&\tfrac{1}{4}d_{abc}d_{abd}\,.\label{dttt}
\eeqa
In obtaining \eqs{fttt}{dttt}, we used the fact that $d_{abc}$ is symmetric and $f_{abc}$ is antisymmetric under the interchange of any pair of indices, which implies that 
\beq \label{fd}
f_{abc}d_{abd}=0\,.
\eeq
To evaluate the products $f_{abc}f_{abd}$ and $d_{abc}d_{abd}$, we proceed as follows.  Using \eqs{commutator}{anticomm},
\beqa 
f_{abd}\Tr(T^a T^b T^c)&=& -i\Tr\bigl([T^b\,,\,T^d]T^b T^c\bigr)=-i\Tr(T^b T^d T^b T^c)+i\Tr(T^d T^b T^b T^c)\,,\\[6pt]
d_{abd}\Tr(T^a T^b T^c)&=& \Tr\left[\left(\{T^b\,,\,T^d\}-\frac{1}{N}\delta_{bd}\mathds{1}\right)T^b T^c\right] \nonumber \\
&=&\Tr(T^b T^d T^b T^c)+\Tr(T^d T^b T^b T^c)-\frac{1}{N}\Tr(T^d T^c)\,.
\eeqa

The traces are easily evaluated using \eqst{trtt}{cf} and (\ref{tatbtatc}), and we end up with
\beqa 
f_{abd}\Tr(T^a T^b T^c)&=&\tfrac{1}{4}iN\delta_{cd}\,,\label{ftr} \\[6pt]
d_{abd}\Tr(T^a T^b T^c)&=&\left(\frac{N^2-4}{4N}\right)\delta_{cd}\,.\label{dtr}
\eeqa
Comparing \eqs{ftr}{dtr} with \eqs{fttt}{dttt}, we conclude that,\footnote{Note that \eqss{fd}{ff}{dd} are equivalent to \eqs{fafa}{fada}, respectively.} 
\beqa
f_{abc}f_{abd}&=&N\delta_{cd}\,,\label{ff} \\[6pt]
d_{abc} d_{abd}&=&\left(\frac{N^2-4}{N}\right)\delta_{cd}\,.\label{dd} 
\eeqa

Consider a $d$-dimensional irreducible representation, $R^a$ of the generators of $\mathfrak{su}(N)$.
The cubic Casimir operator $C_3\equiv d_{abc}R^a R^b R^c$,  commutes with all the $\mathfrak{su}(N)$ generators.  Hence in light of Schur's lemma,  $C_3$ is proportional to the $d\times d$ identity matrix.  In particular, the cubic Casimir operator in the defining representation of $\mathfrak{su}(N)$ is given by
\beq \label{dttt3}
d_{abc}T^a T^b T^c=C_{3F}\mathds{1}\,.
\eeq
To evaluate $C_{3F}$, we multiply \eq{tatb} $d_{abd}$ to obtain
\beq \label{dttid}
d_{abc}T^a T^b=\frac{N^2-4}{2N} T^c\,,
\eeq
after using \eqs{ff}{dd}.  Multiplying the above result by $T^c$ and employing \eq{casimir} yields
\beq \label{dcubic}
d_{abc}T^a T^b T^c=\frac{N^2-4}{2N} C_F\mathds{1}\,.
\eeq
Hence, using \eqs{cf}{dttt3}, we obtain
$$
C_{3F}=\frac{(N^2-1)(N^2-4)}{4N^2} \,.
$$

For completeness, we note the following result that resembles \eq{dttid}, 
$$
f_{abc}T^a T^b=\half\bigl(\{T^a\,,\,T^b\}+[T^a\,,\,T^b]\bigr)=\half f_{abc}[T^a\,,\,T^b]=\half i f_{abc}f_{abd}T^d=\half iNT^c\,,
$$
after employing \eq{ftr}.  Hence, in light of \eqs{casimir}{cf} it follows that
$$
f_{abc}T^a T^b T^c=\half i NC_F\mathds{1}=\tfrac{1}{4} i(N^2-1)\mathds{1}\,.
$$
Indeed, in any irreducible representation of $\mathfrak{su}(N)$, a similar analysis yields
\beq \label{fcubic}
f_{abc}R^a R^b R^c=\half iNC_2\,,
\eeq
where $C_2\equiv R^a R^a$ is the quadratic Casimir operator in representation $R$.
Hence,  $f_{abc}R^a R^b R^c$ is proportional to $C_2$ and thus is not an independent Casimir operator.\footnote{It may seem that \eq{dcubic} implies that the cubic Casimir operator is proportional to the quadratic Casimir operator.  However, the derivation of \eq{dcubic} relies on \eq{tatb}, which only applies to the generators of $\mathfrak{su}(N)$ in the defining representation.   For an arbitrary $d$-dimensional irreducible representation of $\mathfrak{su}(N)$, $C_2$ and $C_3$ are generically independent.}

\section{Matrices of the adjoint representation of SU($N$)}

We now introduce the generators of $\mathfrak{su}(N)$ in the adjoint representation, which will be henceforth denoted by $F^a$.  
The $F^a$ are $(N^2-1)\times (N^2-1)$ antisymmetric matrices, since the dimension of the adjoint representation is equal to the number of generators of $\mathfrak{su}(N)$.  Explicitly, the matrix elements of the adjoint representation generators are determined by the structure constants,
\beq \label{fa}
(F^a)_{bc}=-if_{abc}\,.
\eeq
It is also convenient to define a set of $(N^2-1)\times (N^2-1)$ traceless symmetric matrices 
\beq \label{Da}
(D^a)_{bc}=d_{abc}\,,
\eeq
where the $d_{abc}$ is defined by \eq{dabc}.  Since $d_{abb}=0$ it follows that $\Tr D^a=0$.  The properties of the $F^a$ and $D^a$ matrices have been examined in Refs.~\cite{KR,MacFarlane:1968vc}.

The $F^a$ satisfy the commutation relations of the $\mathfrak{su}(N)$ generators,
\beq \label{Fcomm}
[F^a\,,\,F^b]=if_{abc} F^c\,,
\eeq
which is equivalent to the Jacobi identity,
\beq
f_{abe}f_{ecd}+f_{cbe}f_{aed}+f_{dbe}f_{ace}=0\,.
\eeq
Likewise, there is a second commutation relation of interest,
\beq \label{fdd}
[F^a\,,\,D^b]=[D^a\,,\,F^b]=if_{abc} D^c\,,
\eeq
which is equivalent to the two identities,
\beqa 
&&  f_{abe}d_{cde}+f_{ace}d_{bde}+f_{ade}d_{bce}=0\,, \label{fdiden1} \\
&& f_{abe}d_{cde}+f_{cbe}d_{ade}+f_{dbe}d_{ace}=0\,.\label{fdiden2}
\eeqa
The relations, 
\beq \label{fdd2}
F^a D^b+ F^b D^a=D^a F^b+D^b F^a=d_{abc}F^c\,,
\eeq
are also noteworthy.  Combining \eqs{fdd}{fdd2} yields,
\beq \label{fddf}
F^a D^b+D^a F^b=d_{abc}F^c+if_{abc}D^c\,.
\eeq

The expression for the commutator $[D^a\,,\,D^b]$ is more complicated,
\beq \label{Dcomm}
\bigl[D^a\,,\,D^b\bigr]_{cd}=if_{abe} (F^e)_{cd}-\frac{2}{N}\biggl(\delta_{ac}\delta_{bd}-\delta_{ad}\delta_{bc}\biggr)\,,
\eeq
which is equivalent to the identity,
\beq \label{ffid}
f_{abe}f_{cde}=\frac{2}{N}\biggl(\delta_{ac}\delta_{bd}-\delta_{ad}\delta_{bc}\biggr)+d_{ace}d_{bde}-d_{bce}d_{ade}\,.
\eeq
Interchanging $b\leftrightarrow c$ and subtracting, the resulting expression can be rewritten as
\beq \label{ffpdd}
(F^a F^b+D^a D^b)_{cd}=\frac{2}{N}\biggl(\delta_{ab}\delta_{cd}-\delta_{ac}\delta_{bd}\biggr)+d_{abe}(D^e)_{cd}+if_{abe}(F^e)_{cd}\,.
\eeq
\Eq{ffpdd} is equivalent to the identity,
\beq
f_{ace}f_{bde}-f_{abe}f_{cde}=\frac{2}{N}\biggl(\delta_{ab}\delta_{cd}-\delta_{ac}\delta_{bd}\biggr)+d_{abe}d_{cde}-d_{ace}d_{bde}\,.
\eeq

The quadratic Casimir operator in the adjoint representation is
\beq \label{fafa}
F^a F^a =C_A\id\,,\qquad\quad \text{where $C_A=N$,}
\eeq
and $\id$ is the $(N^2-1)\times(N^2-1)$ identity matrix, which is equivalent to \eq{ff}.
Two other similar expressions of interest are
\beq \label{fada}
D^a D^a=\left(\frac{N^2-4}{N}\right)\id\,,\,\qquad\quad F^a D^a=0\,,
\eeq
which are equivalent to \eqs{dd}{fd}, respectively.

Using the above results, we can derive additional identities of interest.  For example,
\beqa
f_{abc}F^b F^c&=&\half f_{abc}\bigl[F^b\,,\,F^c]=\half i f_{abc} f_{bcd}F^d=\half iN F^a\,, \\[6pt]
f_{abc}F^b D^c&=&\half f_{abc}\bigl[F^b\,,\,D^c]=\half i f_{abc} f_{bcd}D^d=\half iN D^a\,, \\[6pt]
f_{abc}D^b D^c&=&\half f_{abc}\bigl[D^b\,,\,D^c]=\half i\left(f_{abc} f_{bcd}-\frac{4}{N}\delta_{ad}\right)F^d=i \left(\frac{N^2-4}{2N}\right)F^a\,.
\eeqa
It then follows that
\beqa
f_{abc}F^a F^b F^c &=&\half i N^2\id\,,\\[6pt]
f_{abc}D^a F^b F^c &=&0\,,\\[6pt]
f_{abc}D^a D^b F^c &=&\half i(N^2-4)\id\,,\\[6pt]
f_{abc}D^a D^b D^c &=&0\,.
\eeqa
For completeness, we quote the analogous identities with $f_{abc}$ replaced by $d_{abc}$.  These identities are proved in Appendix B of these notes.
\beqa
d_{abc}F^b F^c&=&\half N D^a\,, \label{p1}\\[6pt]
d_{abc}F^b D^c&=&\left(\frac{N^2-4}{2N}\right) F^a\,,\label{p2} \\[6pt]
d_{abc}D^b D^c&=& \left(\frac{N^2-12}{2N}\right)D^a\,.\label{p3}
\eeqa
It then follows that
\beqa
d_{abc}F^a F^b F^c &=&0\,,\label{c3}\\[6pt]
d_{abc}D^a F^b F^c &=&\half(N^2-4)\id\,,\\[6pt]
d_{abc}D^a D^b F^c &=&0\,,\\[6pt]
d_{abc}D^a D^b D^c &=&\left(\frac{(N^2-4)(N^2-12)}{2N^2}\right)\id\,.
\eeqa
\Eq{c3} implies that the cubic Casimir operator in the adjoint representation vanishes.
%, i.e., $d_{abc}F^a F^b F^c=0$.
\clearpage

Finally, we quote a number of useful trace identities~\cite{KR,MacFarlane:1968vc,deAzcarraga:1997ya,Fadin:2005zj}.  
\beqa
&&\Tr F^a=\Tr D^a=0\,,\qquad\qquad\qquad\quad\,\,\, \Tr(F^a D^b)=0\,, \label{t1}\\
&& \Tr(F^a F^b)=N\delta_{ab}\,,\qquad\qquad\qquad\qquad \Tr(D^a D^b)=\left(\frac{N^2-4}{N}\right)\,\delta_{ab}\,,\label{t2}\\
&& \Tr(F^a F^b F^c)=\half iNf_{abc}\,,\qquad\qquad\quad\,\,\,\, \Tr(D^a F^b F^c)=\half Nd_{abc}\,, \label{t3}\\
&& \Tr(D^a D^b F^c)=i\left(\frac{N^2-4}{2N}\right)f_{abc}\,,\qquad\Tr(D^a D^b D^c)=\left(\frac{N^2-12}{2N}\right)d_{abc}\,.\label{t4}
\eeqa
Additional identities involving traces of four generators can also be derived. 
Ref.~\cite{Fadin:2005zj} provides the following results,\footnote{In Ref.~\cite{Fadin:2005zj}, the coefficient of $iNd_{abe}f_{cde}$ in \eq{FDDD} is incorrectly given by $\half$.}
\beqa
\Tr(F^a F^b F^c F^d)&=&\delta_{ad}\delta_{bc}+\half(\delta_{ab}\delta_{cd}+\delta_{ac}\delta_{bd})+\tfrac14 N(f_{ade}f_{bce}+d_{ade}d_{bce})\,,\label{four}\\
\Tr(F^a F^b F^c D^d)&=&\tfrac14 iN(d_{ade}f_{bce}-f_{ade}d_{bce})\,,\label{fourf}\\
\Tr(F^a F^b D^c D^d)&=&\half(\delta_{ab}\delta_{cd}-\delta_{ac}\delta_{bd})+\left(\frac{N^2-8}{4N}\right)f_{ade}f_{bce}+\tfrac14 Nd_{ade}d_{bce}\,,\label{FFDD1}\\
\Tr(F^a D^b F^c D^d)&=&-\half(\delta_{ab}\delta_{cd}-\delta_{ac}\delta_{bd})+\tfrac14 N(f_{ade}f_{bce}+d_{ade}d_{bce})\,,\label{FFDD2}\\
\Tr(F^a D^b D^c D^d)&=&\frac{2i}{N}\,f_{ade}d_{bce}+i\left(\frac{N^2-8}{4N}\right)f_{abe}d_{cde}+\tfrac14 iNd_{abe}f_{cde}\,, \label{FDDD}\\
\Tr(D^a D^b D^c D^d)&=&\left(\frac{N^2-4}{N^2}\right)\delta_{ad}\delta_{bc}+\left(\frac{N^2-8}{2N^2}\right)\delta_{ab}\delta_{cd}+\half\delta_{ac}\delta_{bd}+\tfrac14 Nf_{ade}f_{bce}\nonumber \\
&&\qquad +\left(\frac{N^2-16}{4N}\right)d_{ade}d_{bce}-\frac{4}{N}\,d_{abe}d_{cde}\,.\label{fourd}
\eeqa
Alternative expressions for \eqst{four}{fourd} are given in Appendix C~\cite{deAzcarraga:1997ya}.

As a check of \eq{four}, let us set $a=c$ and sum over $a$.   After employing \eqs{ff}{dd} and relabeling $d$ by $c$, we obtain
\beq \label{f4}
\Tr(F^a F^b F^a F^c)=\half N^2\delta_{bc}\,.
\eeq
Alternatively, one can obtain the above result directly by using eqs.~(\ref{ff}), (\ref{fafa}), (\ref{t2}) and (\ref{t3}) to compute  
\beqa
\Tr(F^a F^b F^a F^c)&=&\Tr\bigl((if_{abd}F^d+F^b F^a)F^a F^c\bigr)=if_{abd}\Tr(F^d F^a F^c)+\Tr(F^b F^a F^a F^c) \nonumber \\[6pt]
&=&if_{abd}\bigl(\half iNf_{dac}\bigr)+N^2\delta_{bc}=\half N^2\delta_{bc}\,,
\eeqa
which confirms the result of \eq{f4}.  Similarly, the results of \eqst{fourf}{fourd} can also be checked by multiplication by either a Kronecker delta, $f_{abc}$ or $d_{abc}$ and then employing the trace formulae previously derived.

Various applications of the identities given in this section can be found in a paper by Roger Cutler and Dennis Sivers~\cite{Cutler:1977qm}.   
Indeed, many of these identities are also reproduced in Appendix~B of Ref.~\cite{Cutler:1977qm}, after correcting the latter for some obvious typographical errors.
The identities provided in these notes are sufficient to work out the color factors for scattering process involving quarks and gluons.  Although the color factors should be computed for the case of $N=3$, it is useful to first evaluate the color factors for an SU$(N)$ gauge theory, since these results allow one to identify sets of independent color factors that arise for a given process.

\section{Two additional identities for $N=3$}

Two additional identities, which were first presented in Ref.~\cite{four}, are special to the case of $N=3$ and do not generalize to arbitrary~$N$.  
These identities can be derived from the characteristic equation of a general element of the $\mathfrak{su}(3)$ Lie algebra~\cite{MacFarlane:1968vc,four},
\beqa
\bigl\{F^a\,,\,F^b\bigr\}_{cd}&=&3d_{abe}(D^e)_{cd}+\delta_{ab}\delta_{cd}-\delta_{ac}\delta_{bd}-\delta_{ad}\delta_{bc}\,,\label{a1}\\
\bigl\{D^a\,,\,D^b\bigr\}_{cd}&=&-d_{abe}(D^e)_{cd}+\tfrac{1}{3}\bigl(\delta_{ab}\delta_{cd}+\delta_{ac}\delta_{bd}+\delta_{ad}\delta_{bc}\bigr)\,.\label{a2}
\eeqa
These two identities can be rewritten as
\beqa
3d_{abe}d_{cde}-f_{ace}f_{bde}-f_{ade}f_{bce}&=&\delta_{ac}\delta_{bd}+\delta_{ad}\delta_{bc}-\delta_{ab}\delta_{cd}\,,\\
d_{abe}d_{cde}+d_{ace}d_{bde}+d_{ade}d_{bce}&=&\tfrac{1}{3}\bigl(\delta_{ab}\delta_{cd}+\delta_{ac}\delta_{bd}+\delta_{ad}\delta_{bc}\bigr)\,.
\eeqa
Combining \eqs{Fcomm}{a1} then yields,
\beq \label{a3}
(F^a F^b)_{cd}=\half i f_{abe}(F^e)_{cd}+\tfrac{3}{2}d_{abe}(D^e)_{cd}+\half\bigl(\delta_{ab}\delta_{cd}-\delta_{ac}\delta_{bd}-\delta_{ad}\delta_{bc}\bigr)\,.
\eeq
Likewise, combining \eqs{Dcomm}{a2} yields,
\beq \label{a4} 
(D^a D^b)_{cd}=\half i f_{abe}(F^e)_{cd}-\half d_{abe}(D^e)_{cd}
+\tfrac{1}{6}\bigl(\delta_{ab}\delta_{cd}-\delta_{ac}\delta_{bd}\bigr)
+\half\delta_{ad}\delta_{bc}\,.
\eeq
Note that the sum of \eqs{a3}{a4} yields the $N=3$ version of \eq{ffpdd}.  Unfortunately, there are no separate analogs of \eqs{a3}{a4} for $N\neq 3$.

%\bigskip\bigskip\bigskip\bigskip
\begin{appendices}

\section{\texorpdfstring{Traces of four generators in the defining representation of SU($N$)}{Traces of four generators in the defining representation of SU(N)}}
\renewcommand{\theequation}{A.\arabic{equation}}
\setcounter{equation}{0}
%\centerline{\large \bf APPENDIX A: Traces of four generators in the}
%\vskip 0.05in
%\centerline{\large \bf  defining representation of SU($N$)}
%\bigskip\bigskip

The trace of a product of four generators in the defining representation also involves the symmetric tensor $d_{abc}$ introduced in Section 2.  Applying \eq{tatb} twice, and taking the trace with the help of \eq{trtt} yields
$$
\Tr(T^a T^b T^c T^d)=\frac{1}{4N}\delta_{ab}\delta_{cd}+\tfrac18\bigl(d_{abe}d_{cde}-f_{abe} f_{cde}+if_{abe}d_{cde}+if_{cde}d_{abe}\bigr)\,.
$$
It is convenient to employ \eqs{fdiden2}{ffid} of Section 3 to produce a more symmetric version,
\beqa 
\Tr(T^a T^b T^c T^d)&=&\frac{1}{4N}\bigl(\delta_{ab}\delta_{cd}-\delta_{ac}\delta_{bd}+\delta_{ad}\delta_{bc}\bigr)+\tfrac18\bigl(d_{abe}d_{cde}-d_{ace}d_{bde}+d_{ade}d_{bce}\bigr)\nonumber \\
&& \qquad\qquad\qquad\qquad\qquad\qquad\,\,
+\tfrac18 i\bigl(d_{abe}f_{cde}+d_{ace}f_{bde}+d_{ade}f_{bce}\bigr)\,.\label{trtttt}
\eeqa
A nice check of \eq{trtttt} is to rederive \eq{tatbtatc} by setting $a=c$ and summing over $a$.

%\clearpage

\section{\texorpdfstring{Proof of \eqst{p1}{p3}}{Proof of eqs. (54)-(56)}}
\renewcommand{\theequation}{B.\arabic{equation}}
\setcounter{equation}{0}

%\centerline{\large \bf APPENDIX B: Proof of \eqst{p1}{p3}}
%\bigskip\bigskip

First, we note that \eqst{p1}{p3} are equivalent to the last three trace identifies of \eqs{t3}{t4},
\beqa
\Tr(D^a F^b F^c)&=&d_{ade}(F^d F^e)_{bc}\,,\\
\Tr(D^a D^b F^c)&=&d_{ade}(F^d D^e)_{bc}\,,\\
\Tr(D^a D^b D^c)&=&d_{ade}(D^d D^e)_{bc}\,,
\eeqa
after using \eqs{fa}{Da}.  Multiplying \eq{fddf} on the left by $F^e$ and taking a trace yields
\beq \label{tffd}
\Tr(F^e F^a D^b)=\half N d_{abe}\,,
\eeq
in light of \eqs{t1}{t2}.  Likewise, multiplying \eq{fddf} on the right by $D^e$ and taking a trace yields
\beq
\Tr(F^a D^b D^e)=\frac{i(N^2-4)}{2N}f_{abe}\,.
\eeq
Multiplying \eq{ffpdd} on the right by $(D^f)_{de}$ and taking the trace (by setting $c=e$ and summing over $e$) yields,
\beq
\Tr(F^a F^b D^f+D^a D^b D^f)=\left(\frac{N^2-6}{N}\right)d_{abf}\,.
\eeq
Finally, we use the result of \eq{tffd} to obtain
\beq
\Tr(D^a D^b D^f)=\left(\frac{N^2-12}{2N}\right)d_{abf}\,.
\eeq

%\bigskip\bigskip\bigskip

\section{\texorpdfstring{Traces of adjoint representation matrices revisited}{Traces of adjoint representation matrices revisited}}
\renewcommand{\theequation}{B.\arabic{equation}}
\setcounter{equation}{0}

%\centerline{\large \bf APPENDIX C: Traces of adjoint representation matrices revisited}
%\bigskip\bigskip

The traces of products of four matrices (either $F^a$ or $D^a$) in the adjoint representation are given in \eqst{four}{fourd}.   It is sometime convenient to eliminate the product $f_{ade}f_{bce}$ in favor of $\delta_{ab}$ and $d_{abc}$, etc.,  by using \eq{ffid}.  The following results were obtained in Ref.~\cite{deAzcarraga:1997ya},
\beqa
\Tr(F^a F^b F^c F^d)&=&\delta_{ab}\delta_{cd}+\delta_{ad}\delta_{bc}+\tfrac14 N\bigl(d_{abe}d_{cde}-d_{ace}d_{bde}+d_{ade}d_{bce}\bigr)\,, \nonumber \\
\Tr(F^a F^b F^c D^d)&=&\tfrac14 iN(d_{abe}f_{cde}+f_{abe}d_{cde})\,,\nonumber \\
\Tr(F^a F^b D^c D^d)&=& \left(\frac{N^2-4}{N^2}\right)\bigl(\delta_{ab}\delta_{cd}-\delta_{ac}\delta_{bd}\bigr)+ \left(\frac{N^2-8}{4N}\right)
\bigl(d_{abe}d_{cde}-d_{ace}d_{bde}\bigr)+\tfrac14 Nd_{ade}d_{bce}\,, \nonumber \\
\Tr(F^a D^b F^c D^d)&=& \tfrac14 N\bigl(d_{abe}d_{cde}-d_{ace}d_{bde}+d_{ade}d_{bce}\bigr)\,,\nonumber\\
\Tr(F^a D^b D^c D^d)&=& i \left(\frac{N^2-12}{4N}\right)f_{abe}d_{cde}+\frac{i}{N}\bigl(f_{ade}d_{bce}-f_{ace}d_{bde}\bigr)+\tfrac14 iN d_{abe}f_{cde}\,,\nonumber \\
\Tr(D^a D^b D^c D^d)&=&  \left(\frac{N^2-4}{N^2}\right)\bigl(\delta_{ab}\delta_{cd}+\delta_{ad}\delta_{bc}\bigr)+  \left(\frac{N^2-16}{4N}\right)\bigl(d_{abe}d_{cde}+d_{ade}d_{bce}\bigr)-\tfrac14 Nd_{ace}d_{bde}\,.\nonumber
\eeqa
Note that the second equation above is consistent with \eq{fourf} in light of \eq{fdd}, and the fifth equation above is consistent with \eq{FDDD} in light of \eq{fdiden1}.

\end{appendices}

%\bigskip\bigskip

\section*{Acknowledgments}

I am grateful to Peter Arnold and 
Eugene Kogan, who encouraged me to make these notes more widely available.   This work is supported in part by the U.S. Department of Energy grant
number DE-SC0010107.

\end{document}